\documentclass[10pt,prb,aps,twocolumn]{revtex4-2}
\usepackage{bm}
\usepackage{amssymb}
\usepackage{graphicx}

\begin{document}

\title{Noncoplanar multi-k states in frustrated spinel and kagome magnets}

\author{M. E. Zhitomirsky}
\affiliation{Universit\'e Grenoble-Alpes, CEA, IRIG-PHELIQS, F-38000, Grenoble, France}
\author{M. V. Gvozdikova}
\affiliation{%
Institut Laue-Langevin, CS 20156, F-38042 Grenoble Cedex 9, France}
\author{T. Ziman}
\affiliation{%
Institut Laue-Langevin, CS 20156, F-38042 Grenoble Cedex 9, France}
        
\date{July 29, 2022}

\begin{abstract}
We investigate analytically and numerically the classical ground states of frustrated Heisenberg models 
on pyrochlore and kagome lattices in zero and finite magnetic fields. Each model has a wide region
 in the microscopic parameter space, where the propagation vector is turned to a commensurate
position equal to a half of the reciprocal lattice vector with a nontrivial star.  Within these regions the zero-field ground states 
for both models correspond to noncoplanar triple-$k$ spin configurations. A
universal appearance of the 3-$k$  states can be related to the spin-space dimensionality. A strong magnetic 
field freezes the longitudinal spin component reducing the spin-space dimensionality.
Accordingly, we find transitions into the double-$k$
magnetic structures induced by applied field for both spin models. The predicted transition between 3-$k$ and 2-$k$ states
may explain the hitherto unexplained transitions observed experimentally in cubic spinels 
GeNi$_2$O$_4$ and GeCo$_2$O$_4$ under magnetic field.
\end{abstract}

\maketitle

\section{Introduction}

We dedicate this article to the memory of Igor E. Dzyaloshinskii who pioneered application
of symmetry methods in magnetism. 

Real materials often display complex magnetic 
structures strikingly different from simple collinear spin arrangements of ordinary ferro- and antiferromagnets. 
Investigation of noncollinear magnetic states began more than
sixty years ago \cite{Dzyal57,Yoshimori59,Villain59,Lyons60,Elliott61,Nagamiya62,Dzyal64}.
Nowadays, they receive  a renewed interest due to their intrinsic multiferroicity \cite{Tokura14} 
and potential for spintronic applications \cite{Baltz18}.  

A novel venue for realization and investigation of complex magnetic structures is offered by geometrically
frustrated magnets \cite{FBook}. Competing exchange interactions between magnetic ions residing on frustrated 
lattices give rise to a large degeneracy of  low-energy spin configurations. Weak additional interactions
present in real materials can lift this degeneracy and stabilize a variety of ordered magnetic
structures. These states demonstrate high tunability with respect to applied field and other external perturbations, 
see, for example, \cite{Reimers91,Harris92,Pinettes02,Domenge05,Bergman07,Lee08,Chern08,Janson08,%
Messio11,Sadeghi15,Maryasin16,Rehn16,Essafi16,Hayami17}.

Here we focus on multi-$k$ structures that may appear once the
star of the instability wave vector has several arms. A few examples of such magnetic states have been
established experimentally, {\it e.g}, for neodymium \cite{Zochowski86,Watson96} and
CeAl$_2$ \cite{Forgan90,Schweizer08}, and discussed theoretically
\cite{Bak78,McEwen86,Harris06,Martin08,Okubo11,Hayami14,Liu16}.
Magnetic structures in all these examples have either incommensurate propagation vectors and/or appear in
metallic systems where coupling to itinerant degrees of freedom is important. Here we focus instead on
magnetic insulators and spin structures based on {\it commensurate} ordering wave vectors.
Specifically, our study is motivated by the antiferromagnetically ordered cubic spinels GeNi$_2$O$_4$ and
GeCo$_2$O$_4$ both exhibiting the propagation vector $(\frac{1}{2},\frac{1}{2},\frac{1}{2})$
in the standard cubic notations
\cite{Crawford03,Diaz06,Matsuda08,Fabreges17,Pramanik19,Basu20}. The star of this ordering wave vector
has four arms corresponding to the four principal cubic diagonals.
Hence, various spin configurations ranging from 1-$k$ to 4-$k$ structures are possible and the fundamental question is
to understand which state is realized depending on temperature and magnetic field.
Remarkably, the same ordering wave vector has been also observed for the pyrochlore antiferromagnet
Gd$_2$Ti$_2$O$_7$ \cite{Champion01,Stewart04} and more recently for Tb$_{2+x}$Ti$_{2-x}$O$_{7+y}$ \cite{Taniguchi13}.
Using neutron diffraction measurements alone, it is  difficult to distinguish a multi-$k$ state from 
multiple single-$k$ domains in a sample. As a result, various suggestions of 
1-$k$  \cite{Champion01}, 4-$k$ \cite{Stewart04,Javanparast15}, and 2-$k$ structures \cite{Paddison20} were put forward
for the pyrochlore  Gd$_2$Ti$_2$O$_7$, whereas the spinel GeCo$_2$O$_4$ was argued to have a 1-$k$ 
magnetic state \cite{Fabreges17}.

In our work we study equilibrium magnetic structures of two spinel materials.  Their qualitative difference 
from the pyrochlore Gd$_2$Ti$_2$O$_7$ is explained in Sec.~\ref{sec:WV}. Furthermore, we
find that the corresponding problem is closely related to finding the classical ground states of a kagome
antiferromagnet with third-neighbor interactions, a so called $J_1$--$J_d$ model
\cite{Janson08,Fak12,Jeschke13,Boldrin15,Ling17}.
For sizable exchange coupling across hexagons $J_d$, Sec.~\ref{sec:WV}B, this model supports 
the noncoplanar cuboc states, which correspond to the 3-$k$ spin configurations
based on the star of the propagation vector $(0,\frac{1}{2})$   \cite{Domenge05,Janson08,Messio11}.
A magnetic transition with the same in-plane ordering wave vector was observed for
kagome material Fe$_4$Si$_2$Sn$_7$O$_{16}$, though it  was argued to have
a single-$k$ structure  \cite{Ling17}.

Below, we  develop an analytical and numerical theory of multi-$k$ states for the above mentioned 
spinel and kagome materials. We discuss in  Sec.~\ref{sec:WV} an appropriate form of spin 
Hamiltonians with competing exchange interactions for each type of the materials
and show in Sec.~\ref{sec:RSMFT} that their lowest-energy spin configuration in zero field 
correspond universally to the 3-$k$ magnetic structures.
Our numerical results for $H=0$, partly overlap with previous theoretical studies
\cite{Domenge05,Messio11,Lapa12,Sim18}, however, we give in Sec.~\ref{sec:LandauZ} new 
analytic arguments, which relate appearance of the 3-$k$ state with 
the dimensionality $D=3$ of the spin space, 
irrespective of the microscopic parameters of the system. In Section~\ref{sec:LandauF}, we extend our theory
to finite magnetic fields and predict a phase transition from a 3-$k$ state
into a 2-$k$ spin configuration in a strong field. The presence of such a phase transition is explained by
an effective reduction of the spin space dimensionality  from $D=3$ to $D=2$ due to freezing longitudinal
spin components in magnetic field, thereby suggesting its universal nature.

\section{Ordering wave vector}
\label{sec:WV}

Insulating transition metal oxides can have distant exchange interactions
between magnetic ions \cite{Janson08,Jeschke13}. Accordingly,
we consider a generic Heisenberg Hamiltonian:
\begin{equation}
\hat{\cal H} = \frac{1}{2} \sum_{n,i}\sum_{m,j} J({\bf r}_{ni}-{\bf r}_{mj}) \:
{\bf S}_{ni}\cdot{\bf S}_{mj} \ ,
\label{Hiso}
\end{equation}
where $n$($m$) is a site index in a unit cell and $i$($j$) is a cell index
on a Bravais lattice such that the magnetic ion position is written as
${\bf r}_{ni} = {\bm \rho}_n + {\bf R}_i$. The $1/2$ prefactor  compensates
for double counting of exchange bonds.

The magnetic ordering wave vector for the spin Hamiltonian (\ref{Hiso}) can be determined
by analyzing the static $\bf q$-dependent susceptibility. In the paramagnetic phase
the susceptibility is diagonal in spin indices but retains a matrix form
for the sublattice index $n$:
\begin{equation}
\chi^{\alpha\beta}_{nm}({\bf q}) = \frac{1}{T}\, \langle S^\alpha_{n,\bf q}\, S^\beta_{m,-\bf q}\rangle
=  \chi_{nm}({\bf q})\,\delta_{\alpha\beta}
\end{equation}
 In the random-phase approximation
the inverse susceptibility tensor is expressed as
\begin{equation}
\chi^{-1}_{nm}({\bf q}) = \chi_0^{-1} \delta_{nm} + J_{nm}({\bf q}) \ ,
\label{chiRF}
\end{equation}
where $\chi_0 = S(S+1)/3T$ is the single-spin susceptibility
and
\begin{equation}
J_{nm}({\bf q}) = \sum_j
J({\bf r}_{ni}-{\bf r}_{mj})\, e^{-i{\bf q}({\bf R}_i-{\bf R}_j)}
\label{Jnm}
\end{equation}
is the Fourier transform of the exchange interactions. At the transition
temperature, the  susceptibility diverges at certain ${\bf q} ={\bf Q}$:
$\det|\chi^{-1}({\bf Q})| = 0$. The corresponding
ordering wave vector ${\bf Q}$ is determined by
the lowest eigenvalue of the exchange
matrix (\ref{Jnm}), which we denote as $J_{\rm min}({\bf Q})$.
For two-dimensional isotropic systems, the random-phase theory fails to predict $T_c=0$, but
the search for the lowest eigenvalue of $J_{mn}({\bf q})$ still corresponds to the first step of the
Luttinger-Tisza method of finding the classical ground states \cite{Luttinger46}.
We perform such calculation separately for spinel and kagome models
with an aim to determine typical values of the exchange parameters that are consistent with
the observed ordering wave vectors.

\subsection{Spinel model}

\begin{figure}[tb]
\centerline{\includegraphics[width=0.7\columnwidth]{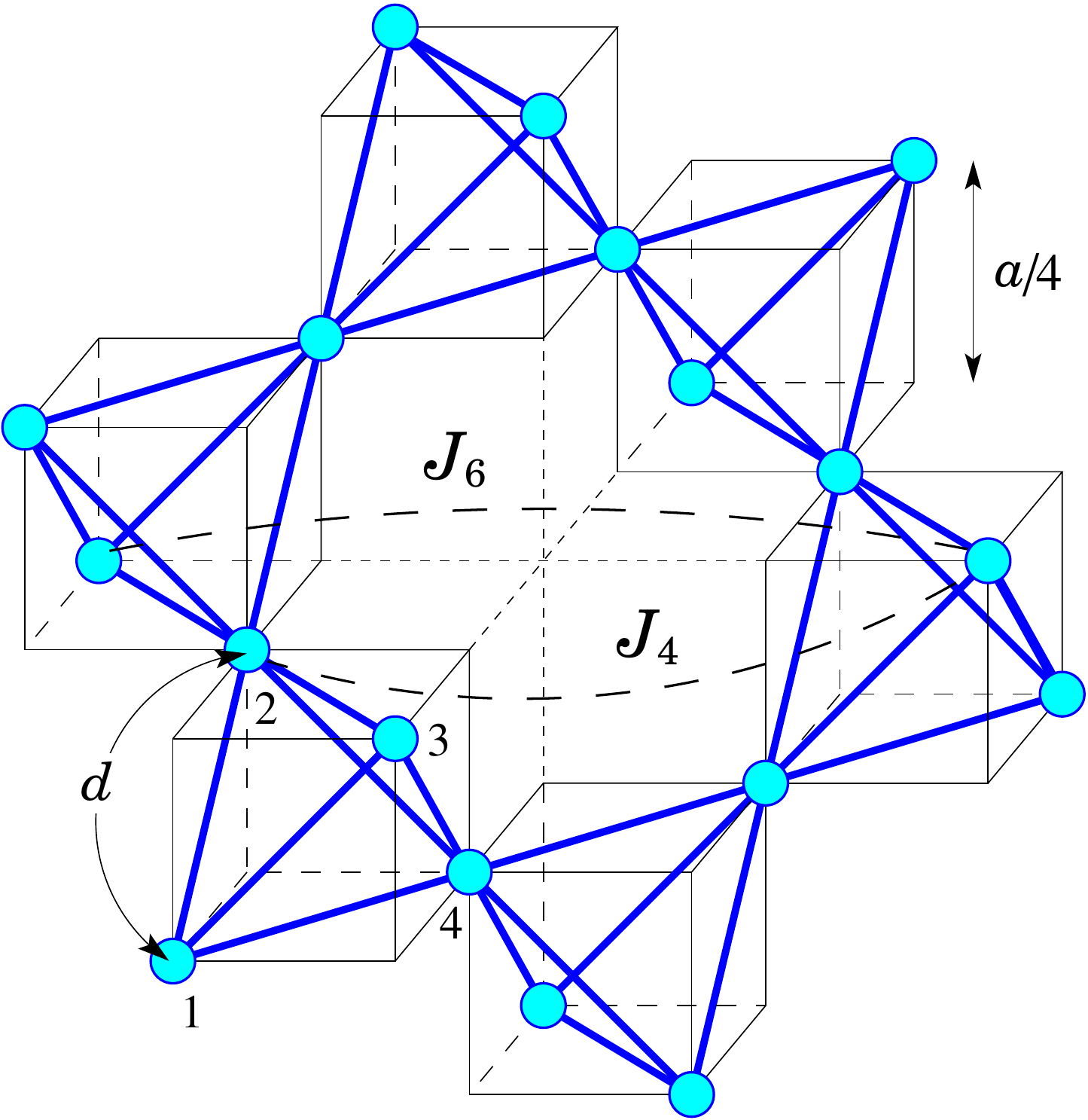}}
\caption{Crystal structure of cubic spinels $AB_2$O$_4$ with B-site cations shown by circles.
The relevant exchanges beyond the nearest-neighbor shell are indicated by long-dashed lines.
Site indices in a unit cell tetrahedron are shown by numbers.
}
\label{fig:spinelL}
\end{figure}

A corner sharing network of tetrahedra formed by magnetic $B^{2+}$ ions in  cubic spinels $AB_2$O$_4$ 
is  presented in Fig.~\ref{fig:spinelL}. Oxygens (not shown) occupy 
the remaining vertices of the small cubes.  A primitive  unit cell contains four magnetic ions located at
${\bm \rho}_1 = (0,0,0)$, ${\bm \rho}_2 = (0,a/4,a/4)$,
${\bm \rho}_3 = (a/4,0,a/4)$, and ${\bm \rho}_4 = (a/4,a/4,0)$.
They form an elementary tetrahedron with the edge $d=a\sqrt{2}/4$, see  Fig.~\ref{fig:spinelL}. 

\begin{figure}[bp]
\centerline{\includegraphics[width=0.6\columnwidth]{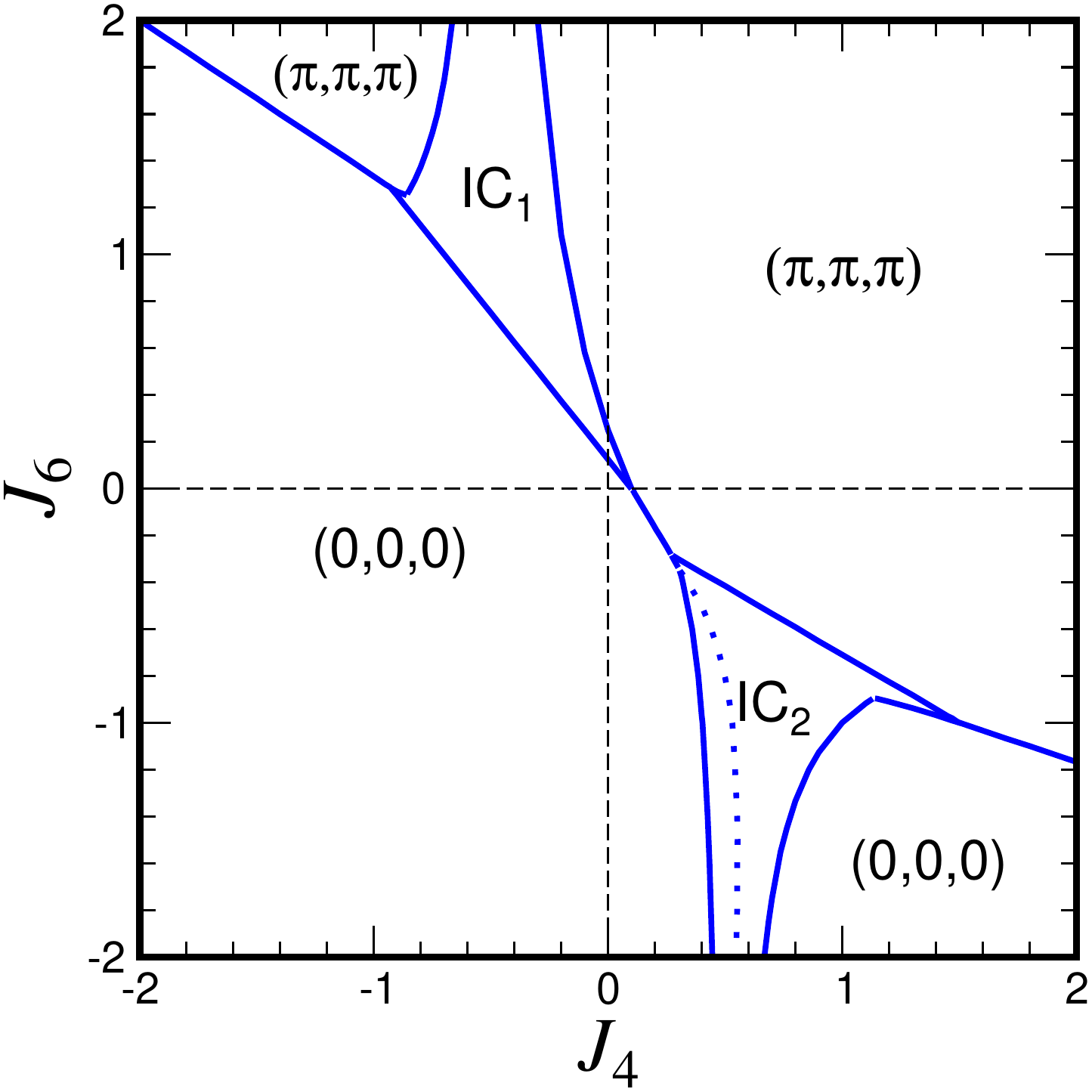}}
\caption{Ordering wave vectors for the $J_1$--$J_4$--$J_6$ spin model
on a pyrochlore lattice with ferromagnetic $J_1=-1$. The IC$_1$ region contains incommensurate 
spirals described by $(q,q,q)$. The IC$_2$ region hosts states with spiral vectors $(q,0,0)$, left part, and $(q,q,0)$,  right part, 
separated by a dotted line.
}
\label{fig:spinelPhD}
\end{figure}

Exchange interactions in GeNi$_2$O$_4$ and GeCo$_2$O$_4$
have been analyzed by Diaz {\it et al.} \cite{Diaz06} using the Goodenough-Kanamori rules.
Here, we briefly summarize their arguments. The nearest-neighbor 90$^\circ$ exchange bonds 
$B$--O--$B$ are ferromagnetic $J_1<0$ in both spinel materials. We shall measure all magnetic 
interactions in units of $|J_1|$ and set $J_1=-1$ where appropriate. The second-neighbor 
exchange for spins at distance $r=d\sqrt{3}$, is also ferromagnetic and, for simplicity, 
can be neglected. The antiferromagnetic couplings between third-neighbors ($r=2d$) across 
hexagons and along chains have similar bond angles and should not differ significantly in 
the spinel structure. We assign a single constant $J_3$ for both of  them. Substantial 
antiferromagnetic couplings are also expected between fourth ($r=d\sqrt{5}$) $J_4$ and 
sixth ($r=2\sqrt{2}d=a$) $J_6$ neighbor spin pairs, see Fig.~\ref{fig:spinelL}. 
In the following, the lattice parameter is set to $a=1$.

We now analyze the effect of antiferromagnetic exchanges couplings
$J_3$, $J_4$, and $J_6$ ($J_1 = -1$) on the stability of
a magnetic order with ${\bf Q} = (\pi,\pi,\pi)$ or
$(\frac{1}{2},\frac{1}{2},\frac{1}{2})$ in the reciprocal lattice units.
The different roles of $J_3$ and $J_6$ are easily understood since they
appear only in the diagonal elements of the exchange matrix:
\begin{eqnarray}
&& J_{nn}({\bf q})=
J_6(\cos q_x+\cos q_y+\cos q_z)
\label{lambda}\\
&& \mbox{}\ +
4J_3 \Bigl(\! \cos\frac{q_x}{2}\cos\frac{q_y}{2} \!+\!
\cos\frac{q_y}{2}\cos\frac{q_z}{2}\!+\!\cos\frac{q_z}{2}\cos\frac{q_x}{2}\!\Bigr).
\nonumber
\end{eqnarray}
For antiferromagnetic $J_6>0$ the six-neighbor contribution has its lowest value for 
${\bf Q}=(\pi,\pi,\pi)$, whereas the third-neighbor contribution vanishes for the same wave 
vector and, hence, cannot stabilize the corresponding antiferromagnetic order 
for either sign of $J_3$. A special role of the six-neighbor exchange in Ge-spinels 
was emphasized  by Diaz {\it et al.}~\cite{Diaz06}, who denoted it as $J_3$.

Results of the numerical search for $J_{\rm min}({\bf q})$ in the plane
$J_4$--$J_6$ ($J_3=0$) are shown in Fig.~\ref{fig:spinelPhD}.
Already quite small $J_{4}^* = 0.1$ and $J_{6}^* = 1/4$ are sufficient to
replace the ferromagnetic state  with an antiferromagnetic structure
described by the propagation wave vector ${\bf Q}=(\pi,\pi,\pi)$.
Thus, the observed magnetic states in GeNi$_2$O$_4$ and GeCo$_2$O$_4$ are stabilized
by exchange interactions between rather distant magnetic ions
from the fourth- and/or the sixth-neighbor shells.
We find in Sec.~\ref{sec:RSMFT} that the equilibrium spin state is
the same in the entire region of the phase diagram Fig.~\ref{fig:spinelPhD}
with the $(\pi,\pi,\pi)$ ordering vector 
irrespective of the ratio $J_4/J_6$. Subsequent
calculations are, then, performed for a simplified $J_1$--$J_6$ ferro/antiferromagnetic model choosing $J_4=0$.
For illustration, we consider two cases: (i) the dominant ferromagnetic interactions  with $J_6/|J_1|=0.5$
and (ii) the dominant antiferromagnetic interactions with $J_6/|J_1|=2$,
for which the Curie-Weiss temperatures
$\theta_{CW}$ are, respectively,  positive and negative.
The former case should model GeCo$_2$O$_4$ with
ferromagnetic $\theta_{CW} = +80$~K, whereas the latter choice of parameters corresponds to
GeNi$_2$O$_4$ with antiferromagnetic $\theta_{CW} = -15$~K \cite{Diaz06}.

\subsection{Kagome model}

The nearest-neighbor Heisenberg antiferromagnet on a kagome lattice has extensive degeneracy
of the  classical ground states. For any triangular plaquette, three spins are constrained
to form the 120$^\circ$ configuration, but periodicity on the lattice is not fixed. In fact,
the lowest eigenvalue of the exchange matrix does not depend on $\bf q$ for this highly
frustrated spin model. It was recognized early on that further neighbor exchanges can lift
degeneracy for the kagome antiferromagnet \cite{Harris92}. In particular, the second neighbor
exchange selects either the $q=0$  ($J_2>0$) or the $\sqrt{3}\times\sqrt{3}$ ($J_2<0$)
coplanar magnetic structures, which correspond to the $\Gamma$ and to the $K$ points in
the Brillouin zone, respectively.

\begin{figure}[t]
\centerline{\includegraphics[width=0.95\columnwidth]{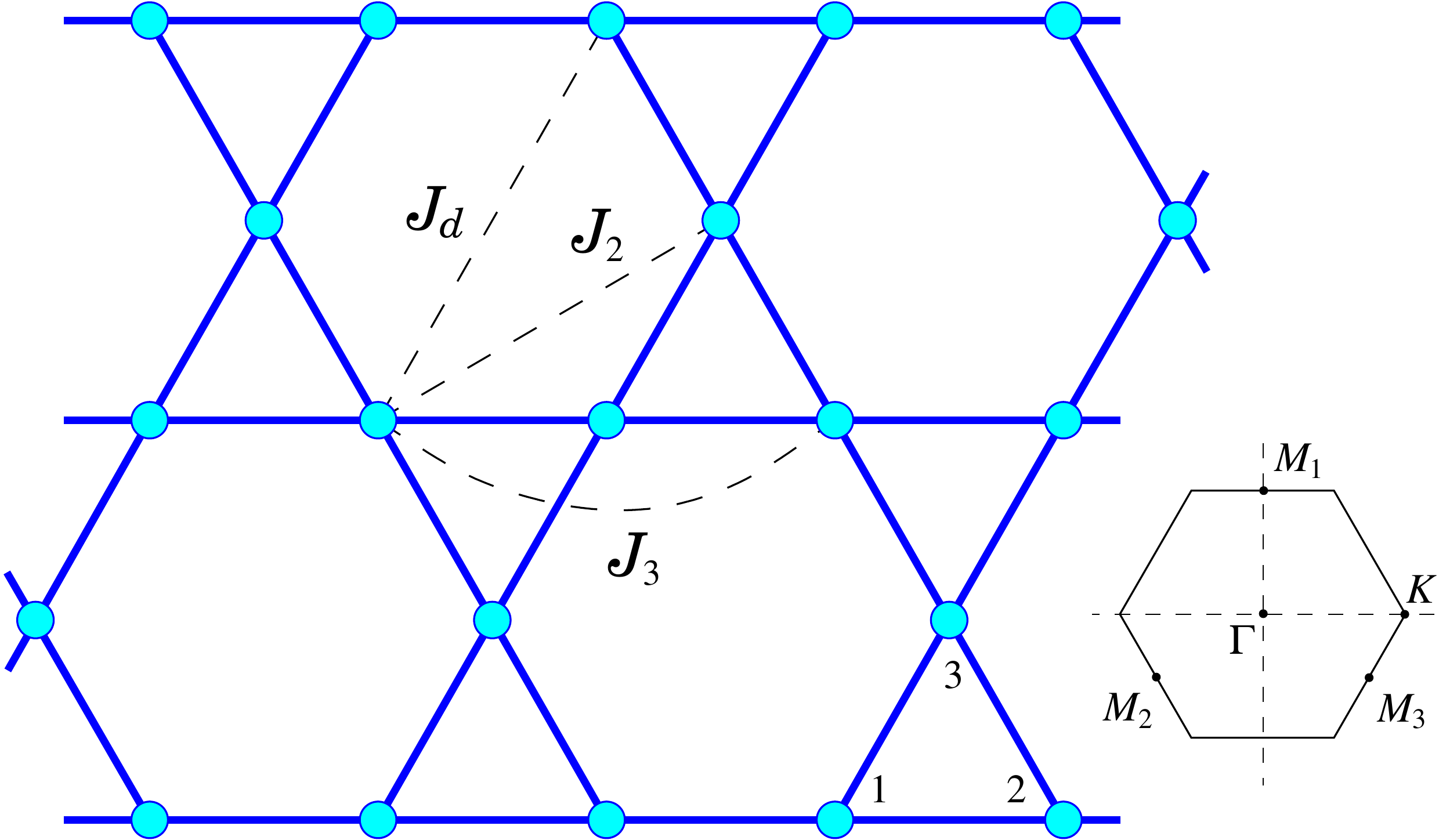}}
\caption{Kagome lattice with  further-neighbor exchanges. Inset:
the hexagonal Brillouin zone with high-symmetry points. 
}
\label{fig:kagomeL}
\end{figure}

\begin{figure}[b]
\centerline{\includegraphics[width=0.6\columnwidth]{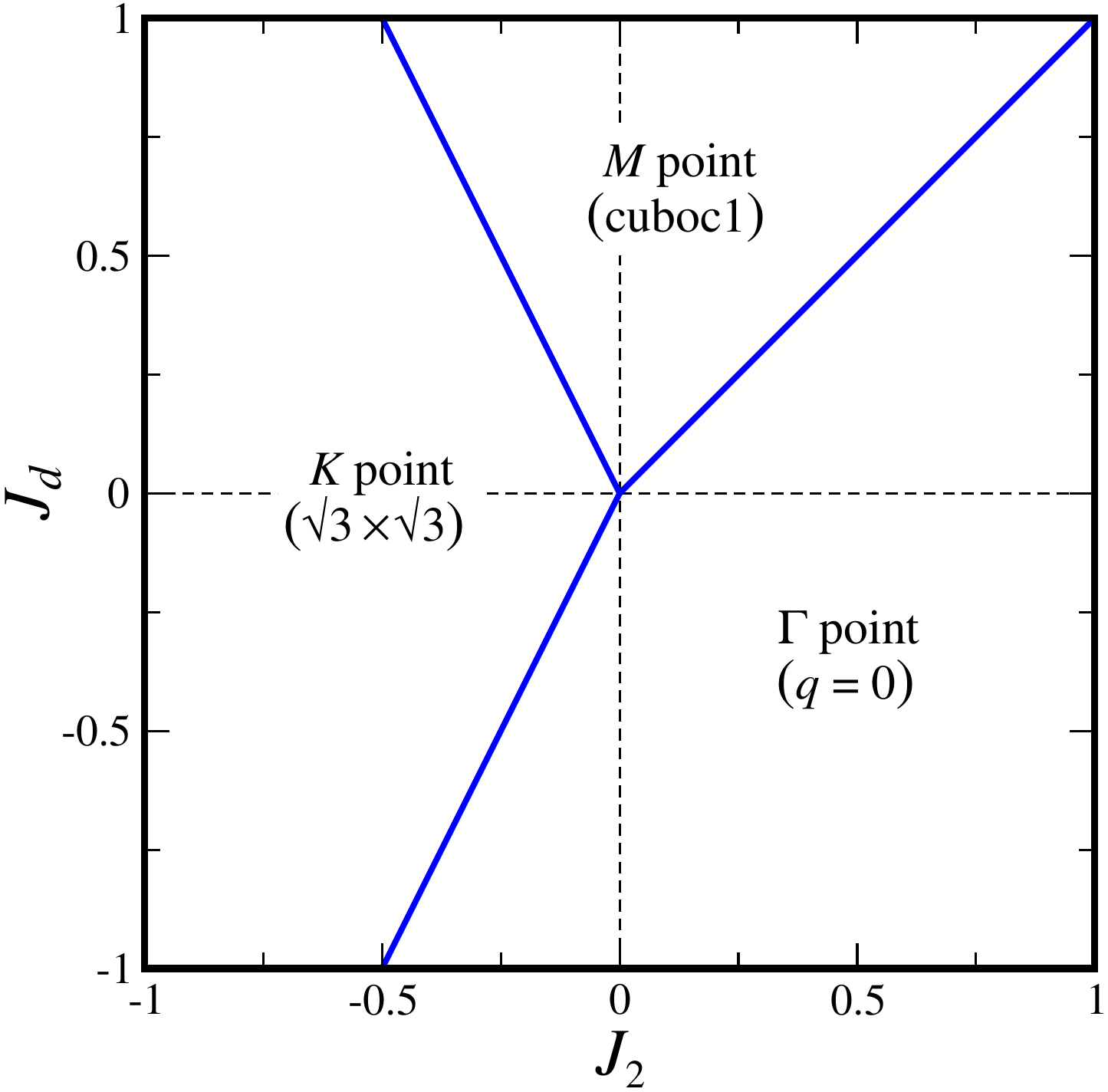}}
\caption{Ordering wave vectors  of the $J_1$--$J_2$--$J_d$ spin model 
on a kagome lattice with antiferromagnetic $J_1=1$ and $J_3=0$.
}
\label{fig:kagome}
\end{figure}

Similarly to the spinel structure, one can identify two types of third-neighbor exchanges,
which have distinct values for the kagome magnets \cite{Janson08,Jeschke13}. We follow
the common notations  and denote them $J_d$
and $J_3$, see Fig.~\ref{fig:kagome}. The effect of $J_3$ on the degeneracy lifting
is basically the same as for $J_2$. It stabilizes the $q=0$ state for $J_3<0$ and
the $\sqrt{3}\times\sqrt{3}$ state for $J_3>0$. The presence of a weak antiferromagnetic
exchange $J_d>0$ has a more interesting consequence producing a noncoplanar spin configuration \cite{Janson08}.
Spins around each triangular plaquette preserve the 120$^\circ$ configuration forming period-2
patterns along the chain directions. In total, there are 12 sublattices that are oriented towards
the corners of a cuboctahedron giving the name cuboc1 to this state \cite{Messio11}.
A similar state called cuboc2 has a 60$^\circ$ angle between neighboring spins
and becomes stable for ferromagnetic $J_1<0$ with large $J_2$
\cite{Domenge05}.

Both cuboc1 and cuboc2 states are based on the star of the wave vector
${\bf Q}_{M_1} = (0, 2\pi/\sqrt{3})$ formed by three edge mid-points of the hexagonal Brillouin zone  
(${\bf Q}_{M} = -{\bf Q}_{M}$), see Fig.~\ref{fig:kagomeL}.
The  wave vector selection for competing $J_2$ and $J_d$ exchanges  ($J_3=0$) is  shown in the lower panel
of Fig.~\ref{fig:kagome}. We have verified numerically that no other states (wave vectors) may intervene
in the shown range of parameters.
Thus, the minimal Heisenberg Hamiltonian for Fe$_4$Si$_2$Sn$_7$O$_{16}$ includes only two 
relevant exchange parameters $J_1$ and $J_d$.

\section{Real-space mean-field theory}
\label{sec:RSMFT}

A common technique for finding the classical ground states of a general exchange Hamiltonian
(\ref{Hiso}) is the Luttinger-Tisza method \cite{Luttinger46}.
A drawback of this scheme is that it cannot be straightforwardly generalized
to anisotropic spin models or to finite magnetic fields. In this section
we use a numerical method called  real-space mean-field simulations.
This method is able to determine equilibrium magnetic structures for an arbitrary spin 
Hamiltonian both at $T=0$ and finite $T$.

The real-space mean-field theory works with a lattice Hamiltonian applying the mean-field 
approximation to decouple products of spins on different sites. Even though the mean-field 
approximation is not quantitatively accurate for nearest-neighbor spin models, it
is usually sufficient to describe  ordered phases at zero and finite temperatures
and to determine topology of the phase diagram. Real-space mean-field
simulations are especially suitable for magnetic systems with competing interactions,
when multiplicity and structure of ordered states are not known in advance
\cite{Bak80,Suzuki83,Cepas04,Gvozdikova16}.

To simplify notations we shall use throughout this Section ${\bf r}_{ni}\rightarrow {\bf r}_i$ 
and $J({\bf r}_{ni}-{\bf r}_{mj})\to J_{ij}$. Two standard steps of the mean-field approximation 
include (i) decoupling of the exchange term
\begin{equation}
{\bf S}_i \cdot {\bf S}_j \approx
{\bf S}_i\cdot{\bf m}_j+{\bf m}_i\cdot{\bf S}_j-{\bf m}_i\cdot{\bf m}_j \ ,
\end{equation}
where  ${\bf m}_i= \langle{\bf S}_i\rangle$,
and (ii) expressing  the spin Hamiltonian (\ref{Hiso})
as
\begin{equation}
\hat{\cal H}_{\rm MF} = -\sum_{\langle ij\rangle}J_{ij}\:{\bf m}_i\cdot{\bf m}_j
 -\sum_i {\bf h}_i\cdot{\bf S}_i \ ,
\end{equation}
with a local field
\begin{equation}
{\bf h}_i = {\bf H} - \sum_j J_{ij}\,{\bf m}_j \ .
\label{Hloc}
\end{equation}
The induced moments for three-component classical vectors
$|{\bf S}_i|=1$ are expressed via the Langevin function:
\begin{equation}
{\bf m}_i = \frac{{\bf h}_i}{|{\bf h}_i|}\Bigl(
\coth\frac{h_i}{T}-\frac{T}{h_i} \Bigr),
\label{Mclass}
\end{equation}
whereas the free-energy is given by
\begin{equation}
{\cal F}_{\rm MF} = -\sum_{\langle ij\rangle}J_{ij}\:{\bf m}_i\cdot{\bf m}_j
-T\sum_i\ln {\cal Z}_i
\label{Fmf}
\end{equation}
with
\begin{equation}
{\cal Z}_i = \frac{T}{h_i} \sinh \frac{h_i}{T} \ .
\label{Zclass}
\end{equation}

The above expressions remain valid for
 models with the anisotropic exchange and the dipolar interactions.
In all these cases,  the ordered moments ${\bf m}_i$ are collinear
with local magnetic fields ${\bf h}_i$.  (The single-ion anisotropy breaks
such a collinearity.)

In quantum case, the ordered component of magnetic moments is expressed via 
 the Brillouin function 
\begin{equation}
{\bf m}_i = \frac{{\bf h}_i}{|{\bf h}_i|} \Bigl[
\Bigl(S+\frac{1}{2}\Bigr)\coth\frac{h_i(S+\frac{1}{2})}{T}
-\frac{1}{2}\coth\frac{h_i}{2T} \Bigr],
\label{Mq}
\end{equation}
whereas the single-spin partition function is
\begin{equation}
{\cal Z}_i = \frac{\sinh[h_i(S+\frac{1}{2})/T]}{\sinh[h_i/2T]} \,.
\label{Zq}
\end{equation}

The mean-field equations are solved iteratively on finite lattices
with linear size $L$ and periodic boundary conditions. These contain
$N=4L^3$ spins for the three-dimensional spinel model and  $N=3L^2$
spins for the kagome lattice. The choice of the ordering wave vectors
for the two models suggests that it is sufficient to have $L=2$ in both cases,
though simulations with $L>2$ have  occasionally been performed as well.

We adopt the following numerical procedure. A random initial configuration
$\{{\bf m}_i^{(0)}\}$ is generated for each set of external parameters and
iterations are performed according to Eqs.~(\ref{Mclass}) or (\ref{Mq}).
The iteration process is stopped once
\begin{equation}
\varepsilon = \frac{1}{N} \sum_{i=1}^{N} |{\bf m}^{(k)}_i -{\bf m}^{(k-1)}_i|
\end{equation}
does not exceed $10^{-6}$--$10^{-7}$. Typically, 20--50 steps
are sufficient to obtain the required accuracy,
though up to $10^3$ iterations may be needed close to phase transitions.
In the end of the iteration process,
various physical quantities are computed including the free energy (\ref{Fmf})
and the magnetic structure factor
\begin{equation}
S^{\alpha\beta}({\bf Q}) = \frac{1}{N}\sum_{i,j}
m^\alpha_{i}  m^\beta_{j} e^{-i{\bf Q}({\bf R}_i-{\bf R}_j)} \,.
\label{Sq}
\end{equation}
In magnetic field, one has to distinguish the longitudinal 
 $S^{zz}$ and transverse components $S^{+-} = S^{xx} + S^{yy}$ of the 
 tensor $S^{\alpha\beta}({\bf Q})$ with respect to the field direction $\bf H\parallel \hat{z}$.

After that, another random spin configuration is generated and the  iteration
process starts over again. Once the second step is finished
the two free energies are compared and the state with a lower ${\cal F}_{\rm MF}$
is kept. Between 10--20 independent iteration runs are performed for each $T$ and $H$
to make sure that the absolute minimum of ${\cal F}_{\rm MF}$ is found.
Such checks are not crucial for the Heisenberg model considered here but may be
important for models  with  single-ion anisotropy \cite{Gvozdikova16}.

\begin{figure}[tb]
\centerline{\includegraphics[width=0.4\columnwidth]{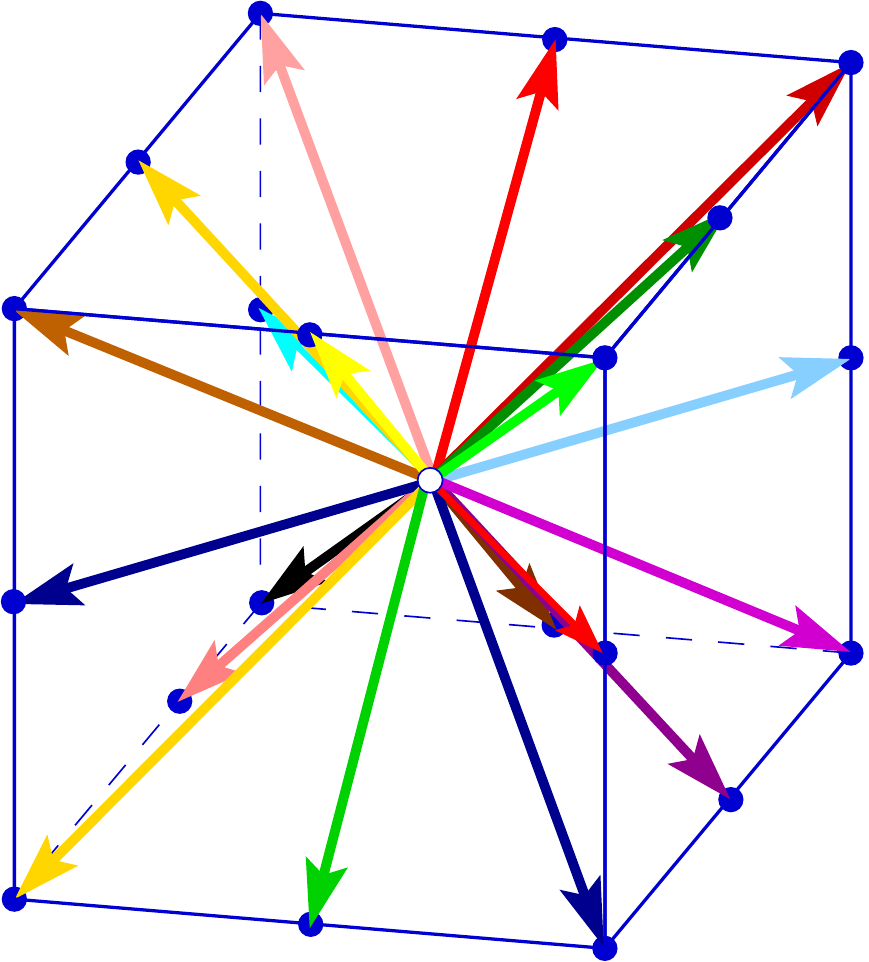}
\hspace*{10mm}
\includegraphics[width=0.4\columnwidth]{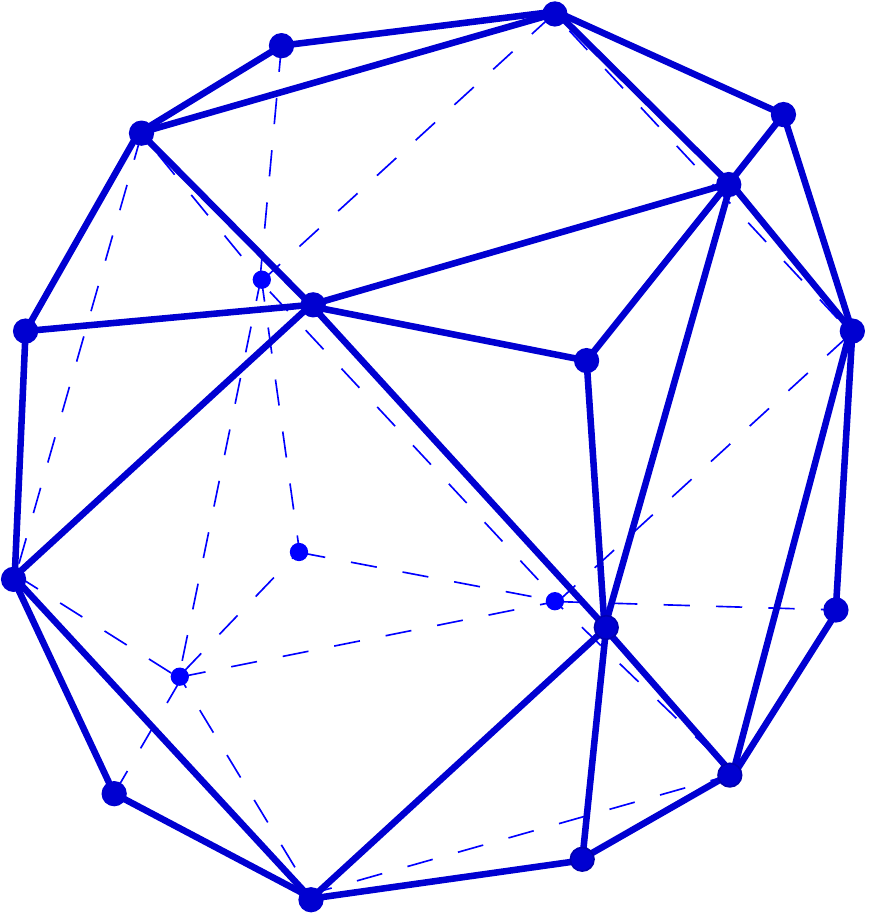}}
\vspace*{3mm}
\caption{Left panel: the common origin plot  of the noncoplanar 3-$k$ magnetic 
structure close to $T_c$. Vectors represent average moments ${\bf m}_i$
for  each of the twenty sublattices.
Right panel: a spherical triacontahedron, which represents the spin structure at $T=0$.
Sublattice moments (not shown) are oriented from the center towards vertices.
}
\label{fig:spinel3k}
\end{figure}

\subsection{Spinel model}

The above numerical  scheme is first applied to investigate the ordered states of 
the $J_1$--$J_6$ spinel model in zero and finite magnetic fields. The qualitative 
behavior is the same for all spin values $S$ and, for illustration, we show only 
results for classical spins. In zero field, there is a single second-order phase 
transition at
\begin{equation}
T_c^{\rm MF} = \frac{1}{3}\,|J_{\rm min}({\bf Q})|
= 2J_6 + \frac{4}{3}\,|J_1| \ .
\end{equation}
Below $T_c^{\rm MF}$ spins form a complex noncoplanar magnetic structure
characterized by three out of four Fourier harmonics belonging to the
star of ${\bf Q} = (\frac{1}{2},\frac{1}{2},\frac{1}{2})$. In total,
this structure has 20 distinct  sublattices distributed over
$4\times 2^3 =32$ sites of the magnetic unit cell. 

A simple visual representation of complex magnetic structures is obtained
with the help of so-called common-origin plots, which draw sublattice
magnetizations in the spin space starting from the same point.
Near $T_c$, an envelope shape of the multisublattice spinel structure is a
cube shown on the left panel in Fig.~\ref{fig:spinel3k}. Twenty sublattices are split into
two groups according to the magnetization length: eight sublattices point towards vertices of 
the cube and have larger magnetization. Twelve remaining sublattices are
oriented towards mid-edge points and have shorter ${\bf m}_i$.
In real space this splitting is arranged in such a way that for every tetrahedron  
one moment is larger than the three others. 

In the mean-field approximation, sublattice magnetizations develop the same length $|{\bf m}_i| = S$
as $T\to 0$. Accordingly, their envelope shape is represented by a polyhedron inscribed in a sphere.
The 3-$k$ structure obtained for spinels corresponds to a 20-vertex polyhedron shown on the right panel 
of Fig.~\ref{fig:spinel3k}. This polyhedron has 30 faces and is accordingly called a 
spherical triacontahedron to distinguish it from other less symmetric polyhedra with 30 faces.
A spherical triacontahedron can be obtained from the commonly known cuboctahedron by attaching 
 pyramids of appropriate height to its triangular faces. As a result, the magnetic structure is highly symmetric
in the spin space with the point group $O_h$. On the other hand, the cubic symmetry of the underlying lattice 
is spontaneously broken due to selection of 3 out of 4 equivalent propagation vectors. An explicit analytic 
form of such a triple-$k$ state, Eq.~(\ref{3KP}), is provided in the following section.

We have verified numerically that precisely the same triacontahedron state is stable 
in the whole region with ${\bf Q} = (\frac{1}{2},\frac{1}{2},\frac{1}{2})$ in Fig.~\ref{fig:spinelPhD}
irrespective of the ratio $J_4/J_6$.
Similar magnetic structures were also obtained for the 
$J_1$--$J_2$ pyrochlore antiferromagnet with both $J$'s positive \cite{Lapa12,Sim18}. 
Furthermore, angles between neighboring spins remain 
the same irrespective of the relative ratios between $J_2$, $J_4$, and $J_6$.  We postpone discussion of a general mechanism
responsible for stabilizing such complex magnetic structure to Sec.~\ref{sec:LandauZ}.

\begin{figure}[bp]
\centerline{\includegraphics[width=0.95\columnwidth]{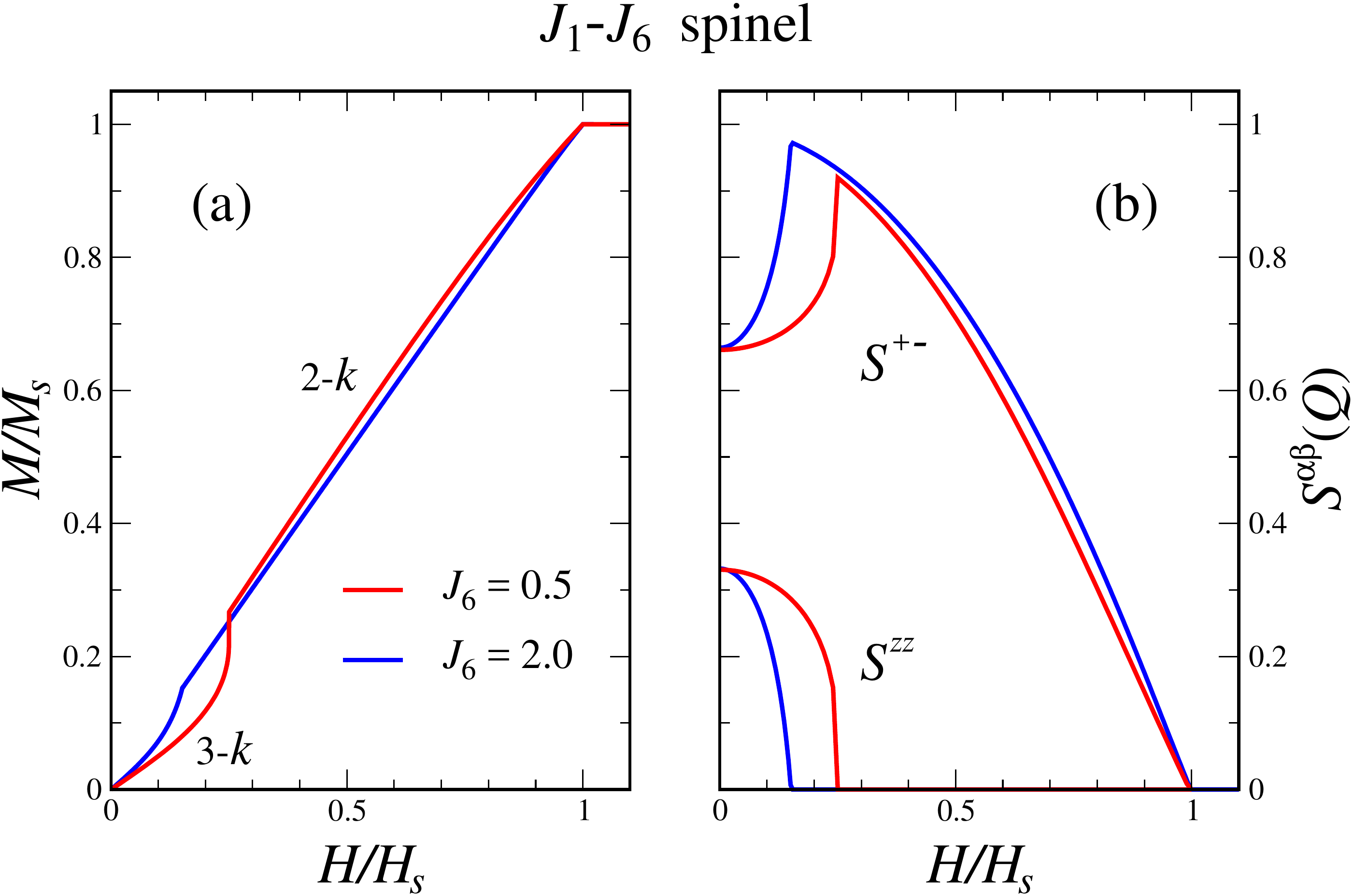}}
\vspace*{3mm}
\caption{Magnetization process of the Heisenberg  $J_1$--$J_6$ spinel antiferromagnet ($J_1=-1$)
at $T=0$  for two values of $J_6/|J_1|$. Left panel: the field-dependence of the uniform magnetization.  
Right panel: the field evolution of the magnetic
structure factor summed over the star of the ordering wave vector $\bf Q$.
}
\label{fig:M_SQS}
\end{figure}

We have simulated the magnetization process at zero and finite temperatures.
Results for $T=0$ are shown in  Fig.~\ref{fig:M_SQS}(a) for two values of $J_6$.
In high magnetic fields spins are oriented parallel to the applied field ${\bf H}\parallel\hat{z}$.
The transition takes place at the saturation field $H_s = 12J_6 -2|J_1|$, see Eq.~(\ref{Hsat}) below.
Remarkably, the magnetization curve of the spinel does not follow the straight line as for majority
of isotropic classical spin models, but
exhibits instead a pronounced upward curvature in the low field regime.
The second-order transition between two antiferromagnetic states takes place at
$H_c/H_s \approx 0.15$ ($J_6 = 2|J_1|$) and  $H_c/H_s \approx 0.25$ ($J_6 = 0.5|J_1|$).
It separates the 3-$k$ spin structure for $H<H_c$ from the 2-$k$ state stable in high fields $H>H_c$.

The field evolution of the magnetic structure factor (\ref{Sq}) summed over all Fourier harmonics
belonging to the star of ${\bf Q}=(\frac{1}{2},\frac{1}{2},\frac{1}{2})$ is shown in  Fig.~\ref{fig:M_SQS}(b).
In the high-field  state, only transverse spin components participate in the
antiferromagnetic ordering, whereas the 3-$k$ state has both $S^{+-}$ and $S^{zz}$
connecting smoothly to the noncoplanar state in zero field.
An interesting aspect of the finite-field behavior is
an orientation of the 3-$k$ spin structure selected by the field for $H<H_c$.
We will discuss  numerical results together with the analytic theory
later in Sec.~\ref{sec:LandauF}. Let us mention here that both cubic spinels
GeNi$_2$O$_4$ and GeCo$_2$O$_4$ exhibit field-induced transitions, which can be related
to the phase transition described above. Such a suggestion is supported by the fact 
that the two materials may have very different local anisotropies \cite{Crawford03,Diaz06}. Hence, it is unlikely 
that these phase transitions derive from a usual competition  between  the anisotropy and the Zeeman energy.

\subsection{Kagome model}

For the $J_1$--$J_d$ kagome model, our real-space mean-field simulations
confirm that the classical ground state in zero magnetic field is the so-called cuboc1
state \cite{Janson08,Messio11}.
This state has 12 sublattices such that relative angles between  nearest-neighbor spins
are equal to 120$^\circ$, whereas all spin pairs connected by $J_d$ bonds are antiparallel.
In the common-origin plot, the twelve sublattices form a cuboctahedron, 
from which the name cuboc was originally derived \cite{Domenge05}.
Such magnetic structure is formed by superposition of all three
Fourier harmonics ${\bf Q}_M$  with equal amplitudes and orthogonal orientation of their spin polarizations.
An explicit analytic expression of this state is given in Sec.~\ref{sec:LandauZ}.
Note that the 3-$k$ state for the kagome antiferromagnet preserves the hexagonal symmetry
of a kagome lattice.

\begin{figure}[tb]
\centerline{\includegraphics[width=0.95\columnwidth]{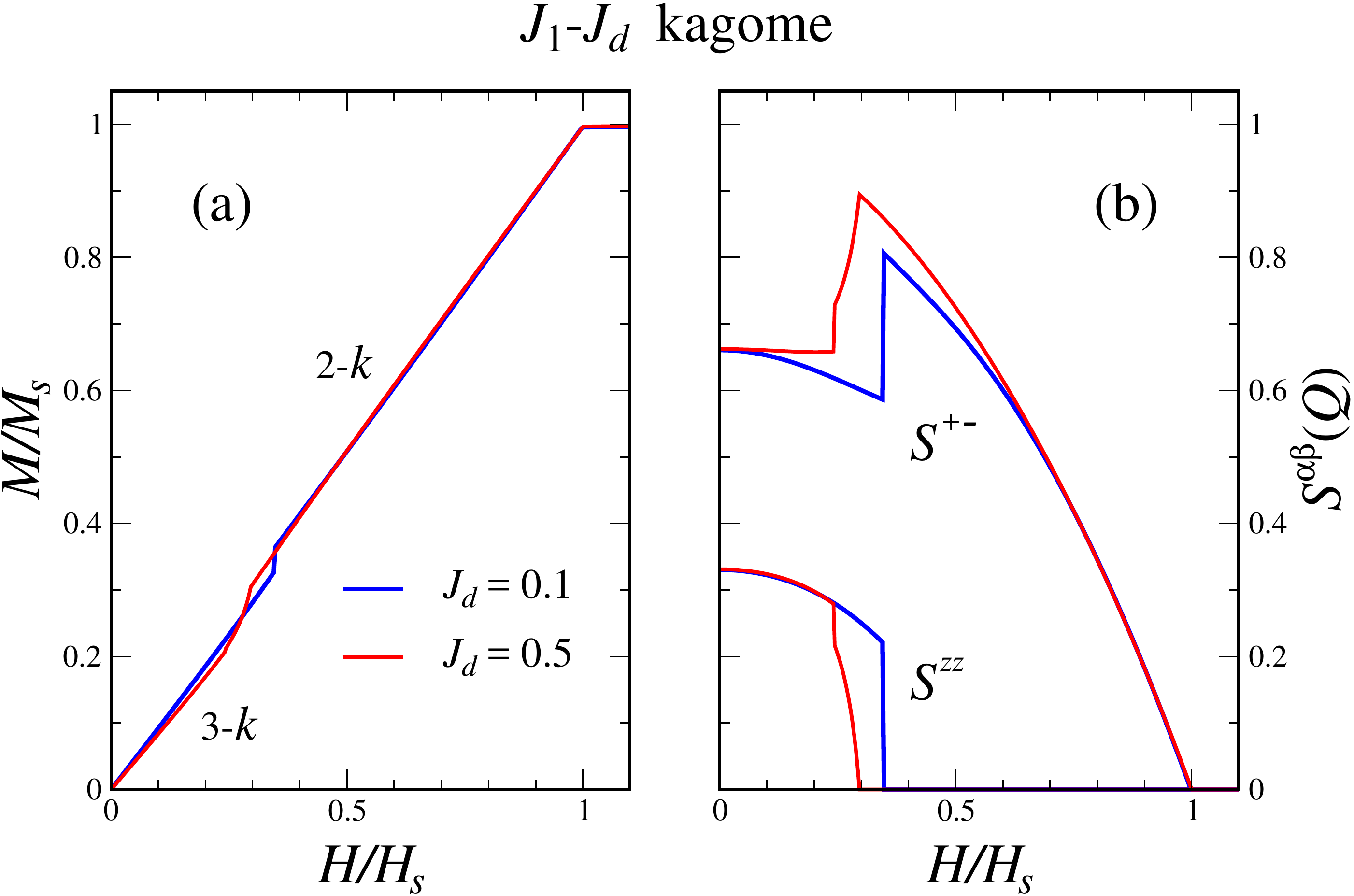}}
\vspace*{3mm}
\caption{Magnetization process of the Heisenberg  $J_1$--$J_d$ kagome antiferromagnet ($J_1=1$)
at $T=0$  for two values of $J_d/J_1$. Left panel: the field-dependence of the uniform magnetization.  
Right panel: the field evolution of the magnetic
structure factor summed over the star of the ordering wave vector $\bf Q$.
}
\label{fig:M_SQK}
\end{figure}

The zero-temperature magnetization curves for the kagome model are shown in Fig.~\ref{fig:M_SQK}(a) for two values of $J_d$.
The saturation field is given by $H_s = 6J_1 + 4J_d$, see Eq.~(\ref{Hsat}).
Deviations from the straight line are significantly smaller than for the spinel model. For $J_d=0.1J_1$
there is a clear first-order jump in the magnetization at $H_c = 0.34H_s$ into a 2-$k$ state.
 Similarly to the spinel model, the high-field state has only transverse components
of the magnetic structure factor,  Fig.~\ref{fig:M_SQK}(b).  However, at low fields the cuboc state
behaves differently by preserving $S^{xx} = S^{yy}= S^{zz}$ or $S^{+-} = 2S^{zz}$ all the way up to $H_c$.

For larger $J_d=0.5J_1$, the field induced transformation of the multi-$k$ state  
becomes a  two step process with transition fields $H_{c1}= 0.24H_s$ and
$H_{c2}= 0.3H_s$ of  first and  second order, respectively. The single transition field $H_c$
splits into a two step transition beginning with $J_d\approx 0.35J_1$.

The above results demonstrate a remarkable similarity between multi-$k$ magnetic structures
stabilized for the isotropic spinel and kagome models in zero and finite magnetic fields.
This similarity points towards a universal mechanism responsible for their selection.
In the next two sections, we present a simple analytic theory explaining the multi-$k$ states formation 
in zero and finite fields.
The remaining differences between the magnetization processes in the low-field region $H<H_c$ appear to be determined
by  symmetry reasons as well.

\section{Landau theory, zero field}
\label{sec:LandauZ}

The Landau expansion for the free-energy is often used to study magnetic systems
close to their transition temperatures. One should distinguish an approach based
purely on symmetry,
which necessarily introduces a large number of unknown phenomenological
parameters \cite{Bak78,McEwen86}, and an energy functional derived from 
a microscopic theory \cite{Reimers91,Plumer90}. In the latter case, the Landau functional
is usually obtained using the density matrix formulation of the mean-field theory. Below,
we briefly sketch an alternative derivation  based on the
$T\rightarrow T_c$ limit of the real-space mean-field theory presented in Sec.~\ref{sec:RSMFT}.
The approach can be straightforwardly generalized to anisotropic models
and to strong magnetic fields, see Sec.~\ref{sec:LandauF}.

In zero applied field, spontaneous moments $m_i=|{\bf m}_i|$
are small and Eqs.~(\ref{Mclass})--(\ref{Zq}) can be expanded near $T_c$ 
in $m_i$ and $h_i$.  Specifically, for classical
spins
\begin{equation}
{\cal F}_i\! =-T\ln {\cal Z}_i \!\approx -\frac{h_i^2}{6T} + \frac{h_i^4}{180T^3}, \ \
m_i \!\approx \frac{h_i}{3T} - \frac{h_i^3}{45T^3}.
\end{equation}
The total free-energy (\ref{Fmf}) is, then, presented as
\begin{equation}
{\cal F}_{\rm MF}= \sum_i \Bigl( \frac{1}{2}\,h_im_i + {\cal F}_i \Bigr) \,.
\end{equation}
Inverting the relation between $m_i$ and $h_i$ and substituting
it back into  ${\cal F}_{\rm MF}$, we obtain the Landau functional as
\begin{equation}
{\cal F}_{\rm MF} \approx \sum_i \Bigl[ \frac{3}{2} T m_i^2
+ \frac{9}{20}Tm_i^4
+ \frac{1}{2}\sum_j J_{ij} {\bf m}_i\cdot{\bf m}_j  \Bigr]\!,
\label{FL0}
\end{equation}
which agrees with \cite{Reimers91}. For quantum spins, the Landau functional has the same form (\ref{FL0})
with additional prefactors $1/S(S+1)$ and $(S^2+S+1/2)/S^3(S+1)^3$ in the
first and in the second term, respectively.

The transition temperature is determined by  $T$, for
which the coefficient in the quadratic term  changes sign from positive
to negative for a certain Fourier harmonic ${\bf m}_{\bf q}$.
It coincides with $T_c$ obtained from the high-temperature
expression for the susceptibility (\ref{chiRF}).
A special feature of the derived functional (\ref{FL0}) is a purely local form of
the quartic term, which does not depend on specific values of $J_{ij}$.
This fact provides significant simplification and universality for the obtained results,
{\it e.g.}, the same noncoplanar state must exist in the entire range of stability of the corresponding
ordering vector.  Furthermore, the results based on Eq.~(\ref{FL0}) remain valid until neglected spin fluctuations
exceed a certain threshold value.

We begin investigation of the stable magnetic structures with the spinel model.
The four ordering wave vectors belonging to the same star are denoted as ${\bf Q}_1 = (\pi,\pi,\pi)$,
${\bf Q}_2 = (-\pi,\pi,\pi)$, ${\bf Q}_3 = (\pi,-\pi,\pi)$, ${\bf Q}_4 = (\pi,\pi,-\pi)$
and the staggered magnetization is represented as
\begin{equation}
{\bf m}_{ni} = \sum_m {\bf l}_m\,\hat{e}_m(n)\,  e^{i{\bf Q}_m{\bf R}_i},
\label{multiK}
\end{equation}
where ${\bf l}_m$ are spin-polarization vectors, which play the role of the order parameter, 
and $\hat{e}_m(n)$ are eigenvectors of the exchange matrix (\ref{chiRF}).
For antiferromagnetic $J_4$ and $J_6$, the  eigenvectors are
\begin{eqnarray}
&& \hat{e}_1 = (0,1,1,1)\,, \ \ \
\hat{e}_2 = (1,0,1,1)\,, \nonumber \\
&& \hat{e}_3 = (1,1,0,1)\,, \ \ \
\hat{e}_4 = (1,1,1,0) \,,
\label{EVectors}
\end{eqnarray}
where the site indexing is shown in Fig.~\ref{fig:spinelL}.
Each $\hat{e}_m(n)$ describes a magnetic structure with parallel alignment
of spins in the kagome planes orthogonal to the corresponding ${\bf Q}_m$ and vanishing moments on
magnetic ions in the intermediate triangular layers. The major difference with the pyrochlore antiferromagnet
Gd$_2$Ti$_2$O$_7$, which orders with the same set of $\{{\bf Q}_m\}$,
is that the eigenvectors (\ref{EVectors}) are nondegenerate and correspond to a one-dimensional
irreducible representation of the small group, whereas the two-dimensional irreducible representation 
is active for Gd$_2$Ti$_2$O$_7$.

Magnitudes and relative orientations of the four vectors
${\bf l}_m$ remain undetermined at quadratic order.
Selection of the equilibrium magnetic structure is produced
by the fourth-order term in the Landau functional (\ref{FL0}).
The minimization procedure is
straightforward for commensurate magnetic structures.
Dropping an unimportant positive prefactor we compute
\begin{equation}
{\cal F}_4 = \frac{1}{N_{\rm mag}} \sum_{i}^{\rm u.c.}\: |{\bf m}_i|^4\,,
\label{F4}
\end{equation}
where summation is performed over a magnetic unit cell,  and minimize ${\cal F}_4$
for a fixed amplitude $l$ of the total order order parameter 
\begin{equation}
\sum_{m=1}^4 {\bf l}_m^2 = l^2\,.
\end{equation}
For the spinel model we have $N_{\rm mag} = 4\times 2^3 = 32$ sites in the magnetic unit cell and
${\cal F}_4$ is explicitly given by
\begin{equation}
{\cal F}_4 =  \frac{1}{4}\! \sum_{m\neq n\neq k}(l_m^2+ l_n^2+ l_k^2)^2
+ 2 \sum_{m\neq n}({\bf l}_m\cdot{\bf l}_n)^2 \ .
\label{F4multi}
\end{equation}
The first term in the above expression favors the simultaneous presence of as many
spin-density harmonics ${\bf l}_m$ as possible, whereas the second term describes their mutual repulsion
once ${\bf l}_m\cdot{\bf l}_n\neq 0$.
Hence, the triple-$k$ structure formed by three orthogonal vectors ${\bf l}_m$ has the lowest
energy ${\cal F}_4^{(3)} = 7l^4/12$ in comparison to the orthogonal 2-$k$ state with
${\cal F}_4^{(2)} = 5l^4/8$ and the 1-$k$ state with ${\cal F}_4^{(1)} = 3l^4/4$.
Once four spin amplitudes ${\bf l}_m$ are present, the repulsive term in Eq.~(\ref{F4multi}) 
always has a nonzero value  and, hence, increases energy for any 4-$k$ state. Assuming a symmetric 
4-$k$ state with a relative angle $\theta = \arccos(-1/3)$
between any pair of ${\bf l}_m$ vectors  we obtain
${\cal F}_4^{(4)} = 31l^4/48$, which is larger than 
${\cal F}_4^{(3)}$ and ${\cal F}_4^{(2)}$. Thus, the number of spin components plays an essential role
in selection of the 3-$k$ state. For an $XY$ model, for example, the energy of a
3-$k$ state increases since the repulsion in the last term cannot be compensated by rotation of vectors
${\bf l}_m$ and the lowest energy would correspond to the 2-$k$ state.

It is instructive to see now how the above conclusion can be reached within a purely phenomenological 
approach. The quartic terms  are given in this case by a sum of three invariants constructed 
from the amplitudes ${\bf l}_m$:
\begin{eqnarray}
{\cal F}_4 & = & \beta_1(l_1^2 + l_2^2 + l_3^2+ l_4^2)^2 + \beta_2 (l_1^4 + l_2^4 + l_3^4 + l_4^4)
\nonumber \\
 & + &\beta_3 \sum_{m\neq n}({\bf l}_m\cdot{\bf l}_n)^2 \,.
\end{eqnarray}
The state selection depends on the signs and magnitudes of $\beta_2$ and $\beta_3$. A comparison with Eq.~(\ref{F4multi}) 
yields $\beta_2>0$ and $\beta_3/\beta_2=2$ in the mean-field theory.
Neglected spin correlations can, in principle, change $\beta$'s. Still, the correlation effects have to be 
sufficiently strong to modify the hierarchy of states based on the energy functional (\ref{FL0}). 
For example, the 1-$k$ state is energetically more favorable than the 3-$k$ state once $\beta_2$ changes the sign, 
whereas the 4-$k$ state becomes lower in energy for $\beta_3/\beta_2<1/2$.

The general form of the 3-$k$ triacontahedron state applicable for all $T\leq T_c$ is
\begin{equation}
\left(\!\begin{array}{c}
{\bf m}_{1i} \\ {\bf m}_{2i} \\ {\bf m}_{3i} \\ {\bf m}_{4i}
\end{array}\!\!\right) = {\bf a} \!
\left(\!\begin{array}{c}
0 \\ 1 \\ 1 \\ p
\end{array}\!\!\right)\! e^{i{\bf Q}_1{\bf R}_i}\! + {\bf b} \!
\left(\!\begin{array}{c}
1 \\ 0 \\ 1 \\  p
\end{array}\!\!\right)\! e^{i{\bf Q}_2{\bf R}_i}\! + {\bf c} \!
\left(\!\begin{array}{r}
1 \\  1\\   0 \\ p
\end{array}\!\!\right)\! e^{i{\bf Q}_3{\bf R}_i}
\label{3KP}
\end{equation}
with ${\bf a}\perp{\bf b}\perp{\bf c}$, $|{\bf a}|=|{\bf b}|=|{\bf c}|$,
and $p\leq 1$. In the vicinity of $T_c$, $p=1$ and one sublattice 
on every tetrahedron has a different length from three others:
$|{\bf m}_4| = \sqrt{3/2}\,|{\bf m}_{1,2,3}|$. On the other hand, at $T=0$
classical spins do not fluctuate and $|{\bf m}_{ni}|=1$, which yields
$p = \sqrt{2/3}$. The classical energy of the triacontahedron state normalized
per one site is
\begin{equation}
E_0/N = - |J_1|\frac{3+2\sqrt{6}}{4} - 3J_6\,.
\end{equation}

Values $p\neq 1$ correspond to admixture of other irreducible 
representations of the small symmetry group of the ordering wave vector
compatible with the residual symmetry of the triacontahedron state.
Hence, the solution (\ref{3KP}) cannot be obtained within the Luttinger-Tisza approach,
which employs only a single  irreducible representation.
Finally, there are eight antiferromagnetic domains of the triacontahedron state
(\ref{3KP}) according to the choice
of the excluded wave vector and the left- or the right-handed triad
$({\bf a},{\bf b},{\bf c})$.

The preceding analysis applies with only slight modifications to the $J_1$--$J_d$ kagome 
model. We denote three propagation vectors as
${\bf Q}_1 = (0,2\pi/\sqrt{3})$ and
${\bf Q}_{2,3} = (\mp\pi,-\pi/\sqrt{3})$, Fig.~\ref{fig:kagomeL}. The respective
 eigenvectors of the exchange matrix are
\begin{equation}
\hat{e}_1 = (1,-1,0)\,,\ \hat{e}_2 = (0,1,-1)\,,\ \hat{e}_3 = (-1,0,1)\,,
\label{EVectorsK}
\end{equation}
where site numbering in a unit triangle goes counterclockwise starting with a lower left  corner,
see Fig.~\ref{fig:kagomeL}.
Similarly to the spinel model, the eigenvector is unique and belongs to
a one-dimensional irreducible representation of the small group of  ${\bf Q}_m$.

Proceeding in a similar way as before we obtain for the kagome model ($N_{\rm mag} = 12$):
\begin{equation}
{\cal F}_4 =  \frac{1}{3} \sum_{m\neq n}\Bigl[(l_m^2+l_n^2)^2
+ 4 ({\bf l}_m\cdot{\bf l}_n)^2\Bigr] \ .
\label{F4multiK}
\end{equation}
The 3-$k$ magnetic structure with orthogonal vectors ${\bf l}_m$ has
${\cal F}_4^{(3)} = 4l^4/9$, which is lower than ${\cal F}_4^{(2)} = l^4/2$
for the 2-$k$ state.

In real space, the triple-$k$ magnetic structure can be represented as
\begin{equation}
\left(\!\begin{array}{c}
{\bf m}_{1i} \\ {\bf m}_{2i} \\ {\bf m}_{3i}
\end{array}\!\!\right)\!\! = {\bf a} \!
\left(\!\!\!\!\begin{array}{r}
 1 \\ -1 \\ 0
\end{array}\!\right)\! e^{i{\bf Q}_1{\bf R}_i}\! + {\bf b} \!
\left(\!\!\!\!\begin{array}{r}
0 \\ 1 \\ -1
\end{array}\!\right)\! e^{i{\bf Q}_2{\bf R}_i}\! + {\bf c} \!
\left(\!\!\!\!\begin{array}{r}
-1\\   0 \\ 1
\end{array}\!\right)\! e^{i{\bf Q}_3{\bf R}_i}\!,
\label{3KK}
\end{equation}
where $({\bf a},{\bf b},{\bf c})$ form an orthonormal triad and
the column index $n=1$--3 corresponds to the site position in a unit cell, Fig.~\ref{fig:kagomeL}.
In contrast to the spinel case, the 12 antiferromagnetic sublattices are equivalent so that
the spin structure (\ref{3KK}) preserves all lattice symmetries gauged by appropriate spin rotations.
Apart from global spin rotations there is only the $Z_2$ degeneracy related to the chirality or left-/right-handedness
of the triad $({\bf a},{\bf b},{\bf c})$. A discrete symmetry breaking in spin models with continuous symmetries
may lead to a finite temperature transition even in two dimensions \cite{Chandra90}.
A finite-temperature transition related to the chiral-symmetry breaking in a Heisenberg kagome
antiferromagnet with the cuboc ground state was demonstrated 
using Monte Carlo simulations by Domenge {\it et al.}~\cite{Domenge08}.

\section{Landau theory, strong fields}
\label{sec:LandauF}

In a finite magnetic field the on-site magnetization ${\bf m}_i =\langle {\bf S}_i\rangle$
is not small. Still, Landau expansion in small transverse components can be used
in the vicinity of the second-order transition line $H_s(T)$ between polarized paramagnetic and 
antiferromagnetic states. Here we focus on the  limit $T\to 0$, $H\sim H_s=H_s(0)$,  
which allows significant simplification for the analytical analysis.

The starting point is the fully polarized state that is stable in high fields $H>H_s$.
Expanding in small transverse spin components $S_i^x \rightarrow m_i^x$, $S_i^y \rightarrow m_i^y$, and
$S_i^z \approx 1 - m_i^2/2 - m_i^4/8$, we obtain for the classical energy
\begin{eqnarray}
E_{\rm MF} & = & \frac{1}{2}\,(H-J_0)\sum_i m_i^2
+ \sum_{\langle ij\rangle} J_{ij}\, {\bf m}_i\cdot{\bf m}_j
\label{Emf}
\\
&  + & \frac{1}{8}\,(H-J_0)\sum_i m_i^4 +
\frac{1}{4}\sum_{\langle ij\rangle} J_{ij}\,m_i^2\,m_j^2\,,
\nonumber
\end{eqnarray}
where $J_0 = \sum_j J_{ij}$ and ${\bf m}_i = (m_i^x,m_i^y)$. The coefficient in front of the quadratic term changes sign at
\begin{equation}
H_s = J_0  - J_{\rm min}({\bf Q}) \ ,
\label{Hsat}
\end{equation}
with $J_{\rm min}({\bf Q})<0$ being the lowest eigenvalue of the exchange matrix (\ref{Jnm}).
The above expression for the saturation field $H_s$ is also valid for quantum spins
with the additional prefactor $S$ on the right-hand side. Thus, below $H_s$ the antiferromagnetic structure is formed by
the same Fourier harmonics of staggered moments
as in the zero-field case (\ref{multiK}) but the polarization
vectors ${\bf u}_m$ are now restricted to lie in the plane perpendicular to the applied field $\bf H$.

Quartic terms in the energy (\ref{Emf}) are represented by two distinct contributions. The first local term
favors the multi-$k$ spin configurations in a full analogy with the zero-field case.
However, since the staggered moments are two-component spin vectors, this term favors  the 2-$k$ magnetic structure,
irrespectively of the choice of $J_{ij}$.
The second quartic term depends explicitly on the exchange parameters and {\it a priori} may select
a different magnetic state.
Performing computations similar to those in Sec.~\ref{sec:LandauZ}
one can compare energies of different multi-$k$ structures.
We have done such an analysis  for the extended  $J_1$--$J_4$--$J_6$ spinel model and the  $J_1$--$J_2$--$J_d$
kagome antiferromagnet and found that the double-$k$ magnetic structures appear in the vicinity of $H_s$
in the whole range of stability of the respective wave vectors in Figs.~\ref{fig:spinelPhD} and \ref{fig:kagome}.
Thus, the field-induced transitions  between the 3-$k$ and the 2-$k$ magnetic structures found numerically in Sec.~\ref{sec:RSMFT}
are completely general in nature and are related to the reduced phase space for staggered moments in a strong magnetic field.

Another aspect of the field behavior of multi-$k$ spin structures is their orientation by
 a weak magnetic field. To address this question from a symmetry prospective we have to return to 
the Landau theory at $H=0$ and add symmetry allowed terms describing interaction of the zero-field order
parameter with an external field $\bf H$. For a general multi-$k$ spin structure one finds  
\begin{equation}
E(H) = \sum_n \Bigl[ \chi_2\,({\bf H}\cdot{\bf l}_n)^2 +  \chi_4({\bf H}\cdot{\bf l}_n)^4\Bigr],
\label{E4H}
\end{equation}
which is valid for both considered models.
 For the orthogonal 3-$k$ magnetic state
the first contribution quadratic in $H$ is isotropic and does not orient the triad $({\bf a},{\bf b},{\bf c})$.
The orientation is determined by the second term. For $\chi_4>0$, it favors a symmetric umbrella-like
orientation of the triad with respect to the field, whereas for $\chi_4<0$ 
one of the three vectors $({\bf a},{\bf b},{\bf c})$ aligns collinearly with the field.
The latter case is consistent with the numerical results of Sec.~\ref{sec:RSMFT}A.
For $0<H<H_c$,  the length of the parallel vector ${\bf l}_n$  is continuously diminished
and eventually vanishes at the second-order transition at $H=H_c$. Accordingly, the longitudinal
structure factor $S^{zz}({\bf Q})$ is gradually suppressed, whereas the transverse components
$S^{xx}({\bf Q}) = S^{yy}({\bf Q})$ somewhat increase towards $H_c$.

The numerical results for the kagome model show a different  behavior in low fields. This difference is explained
by the presence of an additional invariant
\begin{equation}
E_3(H) =  \chi_3 ({\bf H}\cdot{\bf l}_1)({\bf H}\cdot{\bf l}_2)({\bf H}\cdot{\bf l}_3) 
\label{E3H}
\end{equation}
that appears due to ${\bf Q}_1 + {\bf Q}_2 + {\bf Q}_3 =0$ for the kagome model. The interaction term 
$E_3(H)$ scales as $H^3$ and, hence, dominates  over the $H^4$ term for small fields. Depending on 
the sign of $\chi_3$ it selects
either the umbrella ($\chi_3<0$) or the anti-umbrella ($\chi_3>0$) triad orientation relative to $\bf H$.
Both cases correspond to  $S^{xx} = S^{yy}= S^{zz}$ (or  $S^{+-} =2 S^{zz}$) at low fields as seen
 numerically in  Fig.~\ref{fig:M_SQK}(b). Since $S^{zz}=0$ in the field region close to $H_s$, the transition
between low-field and high-field states is of the first order for the kagome model in contrast to the second-order
transition for the spinel model. A  single first-order transition at $H_c$
 changes into a double transition once the two anisotropies $E_3(H)$ and $E_4(H)$ start to compete.
We shall not go into details of such a competition and the corresponding intermediate state at $H_{c1}<H<H_{c2}$
in this work.
\vspace*{3mm}

\section{Discussion}

We have studied the noncoplanar multi-$k$ structures that arise in realistic 
spin models for spinel and kagome magnets. The two considered examples of zero-field states are 
the triacontahedron state with 20 sublattices for the spinel model and 
the cuboc (cuboctahedron) state with 12 sublattice relevant for the kagome materials.
Both magnetic structures are stable in a wide range of microscopic parameters implying a universal
mechanism for their appearance. We argue that such complex magnetic states
are lowest-energy solutions for three-component spins once the star of the commensurate ordering wave vector 
contains several arms. We also demonstrated that an applied field induces 
a phase transition into a 2-$k$ state in high fields in accordance with the spin dimensionality
arguments. The predicted phase transition may explain the experimentally observed
transitions in two spinel materials  GeNi$_2$O$_4$ and GeCo$_2$O$_4$ under magnetic field
\cite{Diaz06,Pramanik19,Basu20}.

An obvious extension of our study is to treat the effect of magnetic anisotropy on  
temperature and field evolution of the multi-$k$ states. Anisotropy has different form 
for the two classes of discussed spin models. The single-ion anisotropy in spinels has 
a staggered local axis parallel to one of the principal cubic diagonals according to
the ion position. For the kagome model, it is more appropriate to include a global easy-plane or easy-axis
anisotropy. A work on these extended models is in progress. In particular, we have found that a 
local easy-plane anisotropy for the spinel model can induce a double transition in zero-field 
in accordance with the experimental results for GeNi$_2$O$_4$ \cite{Crawford03,Basu20}.
An intermediate phase in this case correspond to either a 1-$k$ or a 4-$k$ magnetic structure, whereas
the low-$T$ state retains the shape of a distorted triacontahedron.

Finally, let us mention that noncoplanar magnetic structures provide a novel venue for investigation of
topological magnon bands. The multi-component order parameters identified for the two studied models 
may lead either to new interesting critical behavior or to fluctuation-driven first-order transitions \cite{Chaikin}.

{\bf Acknowledgements.} The work of MEZ was supported by ANR, France, Grant No.\ ANR-15-CE30-0004.

\end{document}